\documentclass[aps,pra,twocolumn,nopacs,superscriptaddress,groupedaddress]{revtex4}

\usepackage{amssymb}
\usepackage{color,graphicx}
\usepackage{amsmath}
\usepackage{amsbsy}
\usepackage{amsthm}
\usepackage{bbm}
\usepackage{bm}
\usepackage{epsfig}
\usepackage{lscape}
\usepackage{float}
\usepackage{subfigure}

\usepackage{wasysym}

\begin{document}

\title{Testing collapse models with levitated nanoparticles: the detection challenge}

\author{A. Vinante$^{\star}$}
\affiliation{Department of Physics and Astronomy, University of Southampton, SO17 1BJ, United Kingdom\\
$\star$ \small{e-mail:  {A.Vinante@soton.ac.uk}}}

\author{A. Pontin$^{\star\star}$}
\affiliation{Department of Physics and Astronomy, University College London, London, WC1E 6BT, United Kingdom\\
$\star\star$ \small{e-mail:  {a.pontin@ucl.ac.uk}}}

\author{M. Rashid}
\affiliation{Department of Physics and Astronomy, University of Southampton, SO17 1BJ, United Kingdom}

\author{M. Toro\v{s}}
\affiliation{Department of Physics and Astronomy, University College London, London, WC1E 6BT, United Kingdom}

\author{P.F. Barker}
\affiliation{Department of Physics and Astronomy, University College London, London, WC1E 6BT, United Kingdom}

\author{H. Ulbricht}
\affiliation{Department of Physics and Astronomy, University of Southampton, SO17 1BJ, United Kingdom}

\date{\today}
\begin{abstract}
We consider a nanoparticle levitated in a Paul trap in ultrahigh cryogenic vacuum, and look for the conditions which allow for a stringent noninterferometric test of spontaneous collapse models. In particular we compare different possible techniques to detect the particle motion. Key conditions which need to be achieved are extremely low residual pressure and the ability to detect the particle at ultralow power. We compare three different detection approaches based respectively on a optical cavity, optical tweezer and a electrical readout, and for each one we assess advantages, drawbacks and technical challenges.

\end{abstract}

\maketitle

\section{Introduction}

Spontaneous wave function collapse (or dynamical reduction) models (CM)~\cite{GRW,CSL1,CSL2,collapse_review1,collapse_review2} have been proposed to reconcile the linear and deterministic evolution of quantum mechanics with the nonlinearity and stochasticity of the measurement process. According to CM, random collapses in space (i.e. localizations) of the wave function of any system occur spontaneously, independently of measurement processes, leading to a progressive spatial localization. The collapse rate scales with the size of the system, leading to rapid localization of any macroscopic system, while giving no measurable effect at the microscopic level, where standard quantum mechanics holds. Importantly, CM lead to a natural solution of the measurement problem, by predicting the emergence of well-defined outcomes in any experiment in agreement with the Born rule.


\begin{figure}[!ht]
\includegraphics[width=9.5cm]{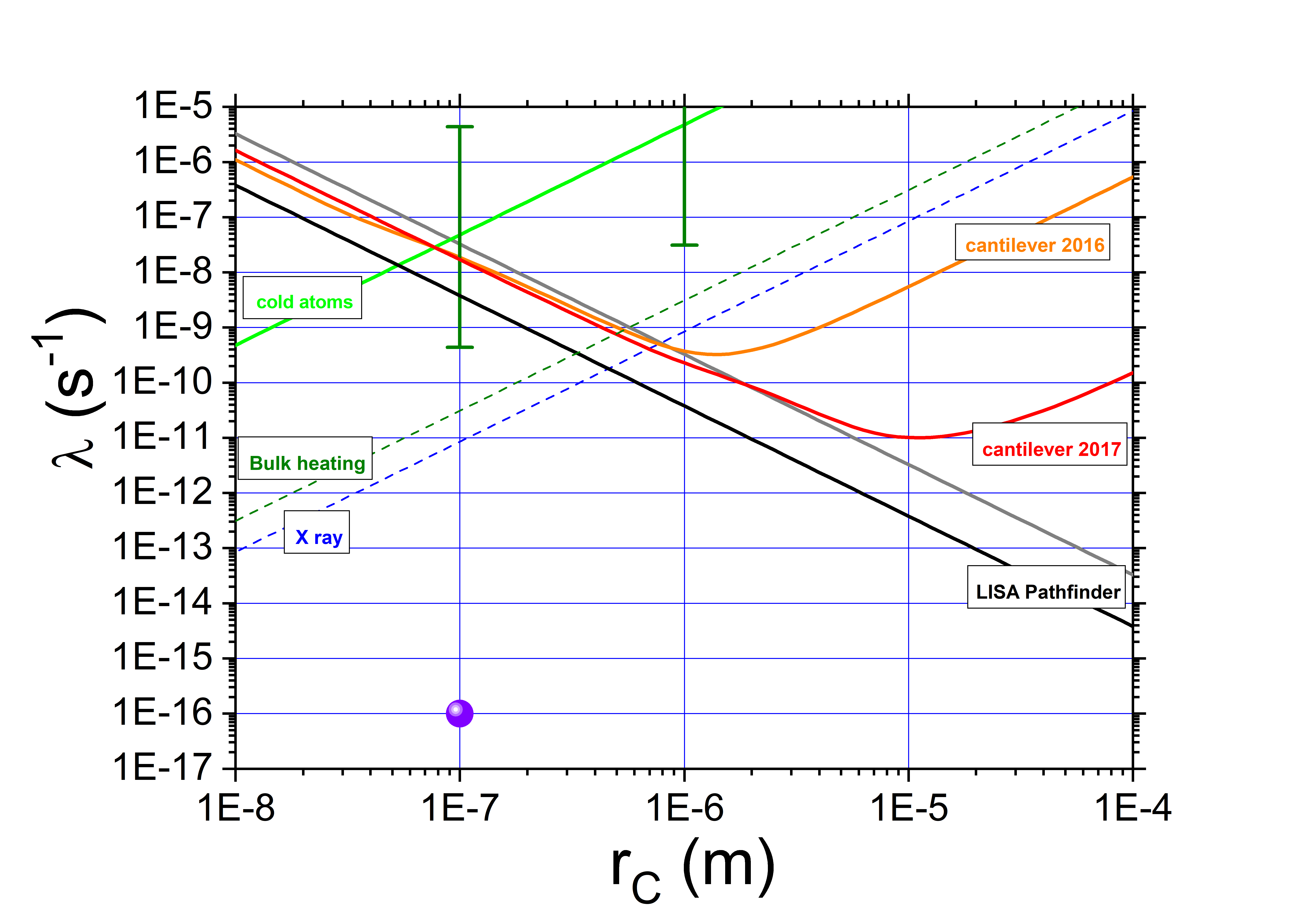}
\caption{Current upper limits on the CSL model collapse rate from different non-interferometric experiments. Solid lines refer to bounds from mechanical or cold atoms experiments, dashed lines to experimental bounds probing the CSL field at very high frequency. Specifically, x-ray spontaneous emission (dashed blue)~\cite{curceanu}, bulk heating in solid matter (dashed dark green)~\cite{adlerbulk}, cold atoms (light green)~\cite{atomi}, cantilever 2016 experiment (orange)~\cite{vinante1}, cantilever 2017 experiment (red)~\cite{vinante2}, LISA Pathfinder early data (gray)~\cite{carlessoLISA,LISA1} and final data (black)~\cite{LISA2}. The theoretical parameters suggested by Adler ~\cite{adler} and GRW~\cite{GRW} are also shown.}
  \label{CSLplot}
\end{figure}

The most general model described in literature is the Continuous Spontaneous Localization (CSL)~\cite{CSL1,CSL2}, a refined version of the earliest collapse model proposed by Ghirardi, Rimini and Weber (GRW)~\cite{GRW}. Other more specific models have been proposed in literature, the most notable being the Di\'{o}si-Penrose gravity-induced collapse model~\cite{penrose1,penrose2,DP}. In this paper we will focus on the CSL model only, as other models can be usually regarded as special cases, or slight variations, of CSL. CSL is characterized by two phenomenological constants, a collapse rate $\lambda$ and a characteristic length $r_C$, which characterize respectively the intensity and the spatial resolution of the spontaneous collapse process. $\lambda$ and $r_C$ are free parameters which should be derived, or bounded, by experiments. The standard conservative values suggested by GRW are $\lambda \simeq  10^{-16}$\,s$^{-1}$ and $r_C=10^{-7}$\,m~\cite{GRW,CSL1} and are sufficient to guarantee almost instantaneous localization of macroscopic objects. A strongly enhanced value for the collapse rate has been suggested by Adler~\cite{adler}, motivated by the requirement of making the wave function collapse effective at the level of latent image formation in photographic process. The values of $\lambda$ suggested by Adler are $\sim  10^{9 \pm 2}$ times larger than the GRW values at $r_C=10^{-7}$\,m, and $\sim  10^{11 \pm 2}$ times larger at $r_C=10^{-6}$\,m.

Several precision experiments have been recently exploited to set significant bounds on the CSL model parameters. The direct effect of collapse models such as CSL is to suppress quantum superpositions, resulting in a loss of coherence in interferometric matter-wave experiments~\cite{expMW}. On top of that, the noise field associated with the collapse implies a violation of the energy conservation. So called noninterferometric tests have been proposed to look for these effects, which include spontaneous emission of x-rays
~\cite{adlerX, curceanu}, force noise in mechanical systems~\cite{collett,adler2005,bassi2005,nimmrichter,diosi, vinante1, vinante2, carlessoLISA, LISA1, LISA2, torsionalCSL,bahrami2014} and spontaneous heating of bulk matter~\cite{adlerbulk} or ultracold atoms~\cite{kasevich, atomi}. At present, noninterferometric tests set by far the strongest bound on CSL parameters, which are summarized in Fig.~\ref{CSLplot}.

Here, we consider noninterferometric tests based on mechanical systems. In this approach, one looks for the universal force noise which is predicted to be induced by CSL in any massive mechanical system. So far, experiments based on ultracold cantilevers~\cite{vinante1,vinante2} and gravitational wave detectors~\cite{carlessoLISA, LISA1, LISA2, torsionalCSL} are setting the strongest bounds. It has been suggested that levitated nano or microparticles would be a nearly ideal platform to perform more sensitive tests~\cite{goldwater,vitali}. Ideally, one needs to work with lowest possible temperature and dissipation in order to minimize thermal noise. We consider a possible experimental implementation based on ion trap techniques, and point out that, for the extremely low thermal noise required by this experiment, the nanoparticle detection becomes a very crucial issue. We consider three possible detection techniques and discuss potential advantages and drawbacks.

\section{Physical system and basic model}

\subsection{CSL noise}


Our goal is to monitor a levitated nanoparticle in order to detect or place strong upper bounds on the universal force noise predicted by spontaneous collapse model such as CSL. The CSL force noise spectral density acting on a homogeneous sphere can be written as~\cite{nimmrichter, diosi, vinante1}:
\begin{equation}
S_{ff,\mathrm{CSL}}= \frac{32 \pi^2 \hbar^2 \lambda r_C^2  \rho^2 R^2}{3 m_0^2} \left[ 1-\frac{2 r_C^2}{R^2}+e^{-\frac{R^2}{r_C^2}} \left( 1+\frac{2 r_C^2}{R^2} \right)   \right]    \label{CSL}
\end{equation}
where $\rho$ is the particle mass density, $R$ is the particle radius and $m_0$ is the nucleon mass.

Note that Eq.\,(\ref{CSL}) grows as $R^2$ for $R \gg r_C$, scaling therefore as a surface noise, and as $R^6$ for $R \ll r_C$. So, if the main background is surface force noise, as for instance due to gas collisions, the signal to noise ratio (SNR) is almost constant for $R>r_C$.
If the main background force noise scales with the volume (as in the case of material-dependent losses) the SNR, according to Eq.\,(\ref{CSL}), will feature a shallow maximum at $R \simeq 2 r_C$. Therefore, if we wish to probe the standard value $r_C=100$\,nm, as a general rule the radius of the particle should be at least of the order of $200$\,nm.

\subsection{Particle}

For simplicity we will consider a spherical nanoparticle, although comparable results could be achieved by other geometries~\cite{torsionalCSL}. We will take SiO$_2$ as standard material. Although it is not the optimal choice for CSL, because of the relatively low density, it is the most used material when optical detection is involved. In general, optical detection requires low absorption dielectric materials, such as silicon or silica. In principle a much larger variety of materials could be used in the case of electrical detection.

\subsection{Paul Trap}

As we wish to work in an ultracold and ultraisolated environment, standard optical trapping is not viable because of the strong heating of the nanoparticle. Alternative levitation methods which avoid this problem are Paul traps~\cite{paultrap} and superconducting magnetic traps~\cite{oriol}.
Both methods cause very small heat dissipation in the levitated particle, so they are expected to be compatible with a cryogenic environment.

We will assume here a Paul trap approach, as related technology has been pushed to quite advanced level by ion trapping community. For sake of simplicity we will not go here into technical details on the geometry of trap. We will only assume that the secular resonant frequency of one relevant translational mode can be set at $f_0=1$\,kHz. Such a value can be obtained by a Paul trap with electrode effective distance of the order of $1$\,mm or smaller, voltage bias of some tens of volts and a charge on the particle of 10-1000\,$e$. These parameters should be also readily compatible with a cryogenic operation.

Among possible sources of noise and decoherence related to the trap, we can mention surface losses in the electrodes and voltage noise in the driving ac and dc bias. Bias voltage noise is a known effect in ion traps, and can be a potentially limiting factor here because the trap potentials have to be kept continuously active. For an order of magnitude estimation, let us assume an electronics voltage noise $S_v=10$\,nV/$\sqrt{\mathrm{Hz}}$ at the secular frequency. This value corresponds to a dynamic range larger than $10^9$, and should be readily achievable, although it seems non-trivial to improve much over this value. Then, for a trapped charge of $q=30$\,$e$, and an effective electrode distance of $500$\,$\mu$m, one estimates a force noise $S_{ff}\simeq  S_v q /d  =9.6 \times 10^{-23}$\,N/$\sqrt{\mathrm{Hz}}$. As shown in Fig.\,\ref{figSfP}, this noise would be comparable with the effect predicted by the CSL model for the standard length parameter $r_C=10^{-7}$\,m and the collapse rate $\lambda=10^{-10}$\,Hz. Therefore, a heavy suppression of electronic noise would be needed in order to probe CSL collapse rate much lower than the latter value. In addition, we note that this noise contribution would increase for larger charge.

\subsection{Environment}  \label {env}
The environment of the particle leads to noise and decoherence through many different channels. A minimal list of sources includes scattering with gas particles and scattering/absorption/emission of thermal photons. In addition, any trapping mechanism typically involves some kind of decoherence. For a Paul trap, besides voltage noise in the driving electrodes, there will be interaction with the electrode surface, as well as electrical losses if an electrical detection circuit is coupled to the trap. Ambient vibrational noise (seismic or acoustic noise) has to be eventually considered.

Collision with gas particles and emission of blackbody radiation will significantly affect the particle dynamics; not only do they represent a noise source but are also the main mechanisms for the thermalization of the particle. So the residual gas pressure $P_g$ and temperature $T_g$, together with the steady state power $W_{\mathrm{abs}}$ absorbed by the particle will determine its equilibrium bulk temperature $T$ as discussed in the following section.


\subsubsection{Thermal equilibrium}
The heat flow from a hot nanoparticle to a cold surrounding gas in the molecular regime can be calculated using the formula~\cite{liu}:
\begin{equation}
\dot{Q}_{gas}=-\frac{\alpha \pi R^2 P_g v_t}{2 T_g}\frac{\gamma_s+1}{\gamma_s-1} \left( T-T_g \right)  \label{conductiontogas}
\end{equation}
where $\alpha$ is a thermal accommodation factor, $v_t=\sqrt{8 k_B T_g/\left( \pi m \right) }$ is the gas thermal velocity, with $m$ molecular mass, and $\gamma_s$ is the specific heat ratio. Here all parameters can be easily determined, except for the accommodation factor $0<\alpha<1$. Typical values of order $0.4$ are reported in literature for specific experimental situations~\cite{liu}.

The heat flow by blackbody radiation is described by the expression~\cite{chang}:
\begin{equation}
\dot{Q}_{bb}=-\frac{72 \zeta(5)}{\pi^2}\frac{V k_B^5}{c^3 \hbar^4} \mathrm{Im}\frac{\epsilon_{\mathrm{bb}}-1}{\epsilon_{\mathrm{bb}}+2} \left( T ^5 - T_g^5 \right ) \label{conductiontoradiation}
\end{equation}

\noindent where $V$ is the particle volume and $\zeta(5)\approx1.04$ is the Riemann zeta function. The dependence on $T^5$ is typical of a subwavelength nanoparticle. Here, the flow is controlled by the absorption coefficient $\epsilon_{\mathrm{abs}}=\mathrm{Im} \left[ \left( \epsilon_{\mathrm{bb}}-1 \right)/ \left( \epsilon_{\mathrm{bb}}+2\right) \right]$ where $\epsilon_{\mathrm{bb}}$ is the blackbody emissivity. For typical situations, as for instance silica at $100$\,K, this term can be taken of the order of $0.1$.

\begin{figure}[!ht]
\includegraphics[width=8.6cm]{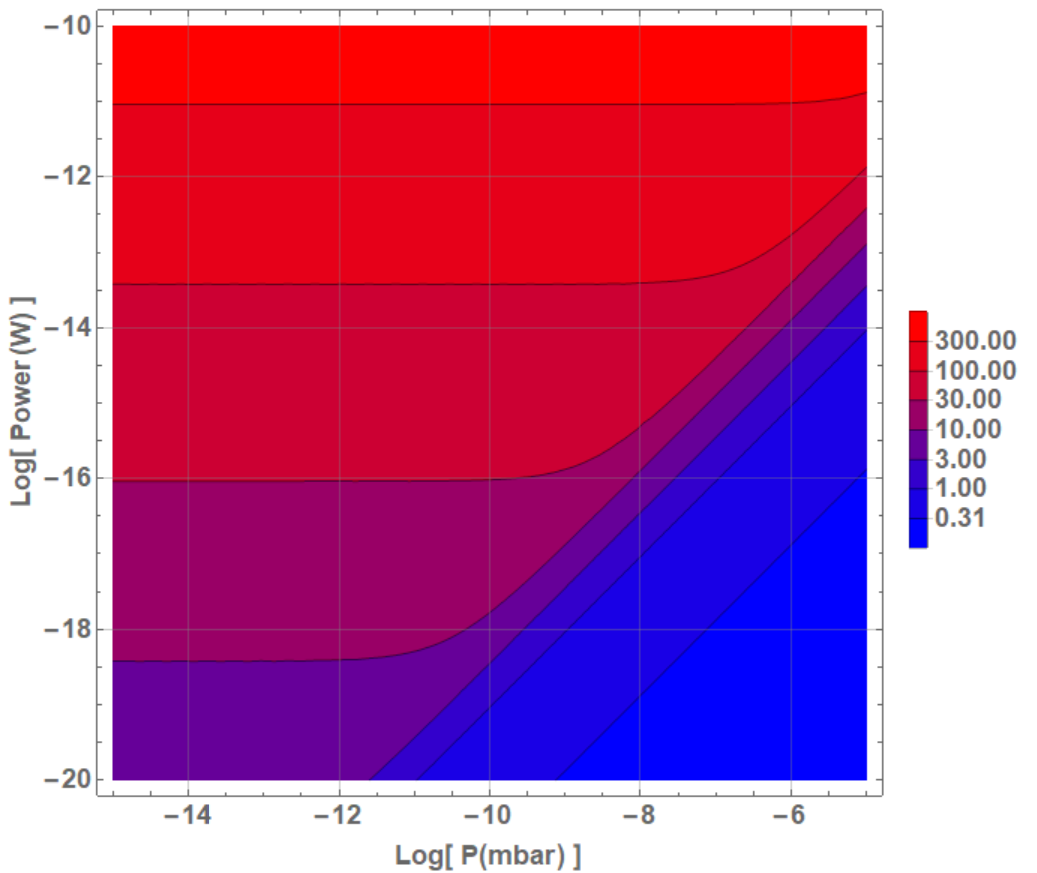}
\caption{Internal temperature of the nanosphere as a function of the gas pressure and the absorbed power. The particle is a silica nanosphere with $R=200$\,nm, the residual gas is helium at $T_g=300$\,mK, the thermal accomodation factor has been set to $\alpha=0.4$ and the emissivity to $\epsilon_{\mathrm{abs}}=0.1$.}
  \label{figTvsP}
\end{figure}

By setting $\dot{Q}_{bb}+\dot{Q}_{gas} + W_{\mathrm{abs}}=0$ one can estimate the equilibrium internal temperature $T$ of the particle as function of the environmental conditions and the input power. Fig.\,\ref{figTvsP} illustrates the dependence of the equilibrium temperature of the nanosphere on the gas pressure $P_g$ and absorbed power $W_{\mathrm{abs}}$ in a relevant region of the parameter space. Two regimes are clearly visible. For low pressure/high power the particle is thermalized by radiation, and the equilibrium temperature is independent of pressure. For high pressure/low power the particle is thermalized by the gas and the equilibrium temperature depends solely on the ratio $W_{\mathrm{abs}}/P_g$.

\subsubsection{Thermal force noise}
To estimate the thermal force noise due to the gas in a hot-particle scenario we follow the model of ref.~\cite{millen}, i.e. we separately consider the contributions due to the impinging and emerging gas particles. The model assumes two different baths with temperature $T_i=T_g$ for the impinging molecules and $T_e=T_i+ a \left( T-T_i \right)$ for the emerging molecules where $a$ is another phenomenological accommodation factor $0<a<1$. The underlying idea is that the scattering of a gas molecule off the nanoparticle is not elastic: the particle is assumed to partially thermalize with the nanoparticle before being re-emitted. This model of scattering is abundantly supported by experimental literature~\cite{millen}.

The two baths lead to different mechanical damping rates:
\begin{align}
&\Gamma_{i}=\frac{4 \pi}{3} \frac{m R^2 v_t P_g}{k_B T_i m_s} \\
&\Gamma_{e}= \frac{\pi}{8} \sqrt{\frac{T_e}{T_i}} \Gamma_{i}
\end{align}
where $m_s$ is the mass of the particle.
The force noise can then be calculated as
\begin{equation}
S_{ff,\mathrm{g}} =4 k_B m_s \left( \Gamma_i T_i+\Gamma_e T_e \right).
\end{equation}

There will also be, even in absence of detection, a force noise due to recoil from emission of blackbody radiation. Following calculations similar to the one leading to Eq.\,(\ref{conductiontogas}), one arrives at the formula~\cite{chang}:
\begin{equation}
S_{ff,\mathrm{bb}}=\frac{160}{\pi}\frac{R^3 k_B^6}{c^5 \hbar^4} \mathrm{Im}\frac{\epsilon_{\mathrm{bb}}-1}{\epsilon_{\mathrm{bb}}+2}  T ^6. \label{bbnoise}
\end{equation}

It is easy to check that this contribution is exceedingly small compared to the one of the gas for any realistic set of parameters, except under the condition of extremely low pressure below $10^{-15}$\,mbar and relatively high power. The reason is that photons can remove efficiently energy but at the same they carry very little momentum. An even smaller contribution arises from the scattering and absorption of blackbody radiation coming from the cold environment at temperature $T_g$.

\begin{figure}[!ht]
\includegraphics[width=8.6cm]{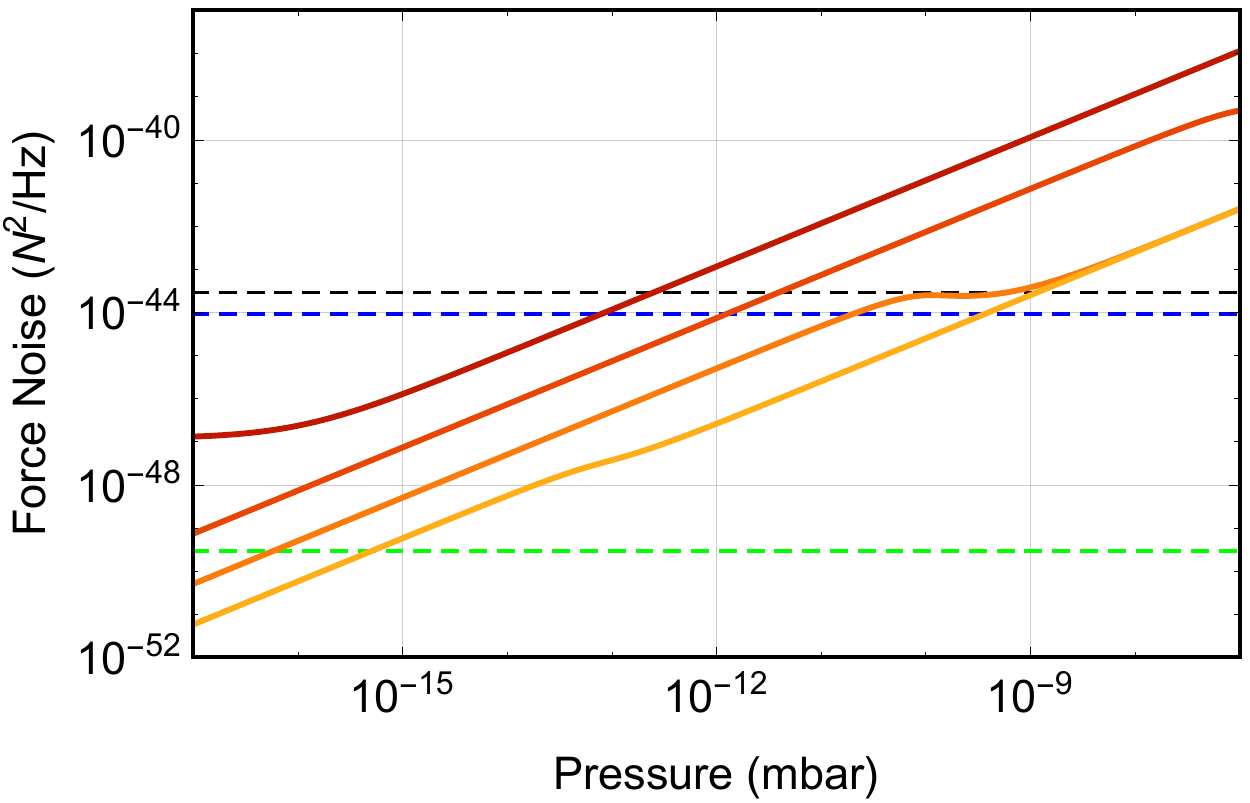}
\caption{Thermal force noise as a function of the residual gas pressure. It includes the effects of both gas collisions and blackbody radiation. Thick solid curves refer from top to bottom to an absorbed power of $10^{-10}$,$10^{-14}$,$10^{-18}$,$10^{-22}$\,W. Blackbody radiation is always negligible except for the flattening of the $10^{-10}$\,W curve at low pressure. As in Fig.\,\ref{figTvsP}, the particle is a silica nanosphere with $R=200$\,nm, the residual gas is helium at $T_g=300$\,mK, the thermal accommodation factor has been set to $\alpha=0.4$ and the blackbody emissivity to $\epsilon_{\mathrm{abs}}=0.1$. As a reference we also plot the estimated contribution from the Paul trap bias noise (middle horizontal blue dashed line), see text for details, and the CSL force noise for $r_c=10^{-7}$\,m, $\lambda=10^{-10}$\,Hz (top dashed black line) and  GRW values $r_c=10^{-7}$\,m, $\lambda=1 \times 10^{-16}$\,Hz (bottom dashed green line). }  \label{figSfP}
\end{figure}

Fig.\,\ref{figSfP} shows the thermal force noise as a function of gas pressure for different values of the absorbed power. Clearly, going to sufficiently low pressure will eventually suppress thermal noise below any detectable level. For instance, for the lower value of the pressure reported in literature $10^{-17}$\,mbar~\cite{gabrielse} and absorbed power $10^{-18}$\,W, the thermal noise would compare to the extremely tiny effect of CSL according to the GRW values. Unfortunately, other effects become dominant in this regime, in particular electrical noise in the Paul trap and backaction noise from the detection.

\section{Detection schemes}
Here we come to the key issue we wish to study in this paper, namely how to choose and to optimize the detection of the particle. We will consider three detection options. An optical cavity readout is discussed in Sec.\,\ref{sec_cav}, an optical tweezer in Sec.\,\ref{sec_tweezer} and an electrical readout based on a SQUID in Sec.\,\ref{sec_squid}. Before going into the details of the three techniques, we will analyze in Sec.\,\ref{bandwidth} the general features of two different measurement strategies, a stationary continuous one and a stroboscopic reheating, finding that they are in principle equivalent. Based on this conclusion we will focus in the following sections on the continuous measurement strategy.

\subsection{Continuous and stroboscopic measurement}\label{bandwidth}

In this section we will try to compare continuous and stroboscopic measurements. Some relevant considerations can be done regardless of the specific detection technique.

In a steady-state approach the position of the trapped mechanical harmonic oscillator is continuously measured. The acquired signal is Fast-Fourier-transformed (FFT) and periodograms are averaged to provide an estimation of the power spectral density (PSD). An example of application of this method in the context of testing collapse models is given by recent cantilever experiments~\cite{vinante2}.

In general, there will be two contributions to the PSD, a wideband position measurement noise $S_{xx}$ and the true oscillator noise with PSD given by $|\chi \left( \omega \right)|^2 S_{ff} $ where:
\begin{equation}
\chi \left( \omega \right)= \frac{1}{m_s \left[ \left( -  \omega^2+w_0^2 \right) -i \omega \Gamma\right]}
\end{equation}
is the Lorentzian mechanical susceptibility. For high-Q systems, and provided $S_{xx}$ is low enough, the oscillator noise will be dominant around the resonant frequency $\omega_0$, over a given bandwidth $\Delta f=\Delta \omega/2 \pi$ which depends on the actual values of $S_{xx}$ and $S_{ff}$. If we define $\Delta f= |f_2 -f_1|$, where $f_{1,2}=\omega_{1,2}/2\pi$ are the frequencies at which $S_{ff}| \chi \left( \omega \right)|^2 = S_{xx}$, we find that:
\begin{equation}
  \Delta{f} = \frac{1}{2 \pi m_s \omega_0} \sqrt{\frac{S_{ff}}{S_{xx}}}   \label{noisek}
\end{equation}

The estimation of $ S_{ff}$ is inferred from the data available within this bandwidth, for instance by fitting the estimated PSD. The relative uncertainty on $S_{ff}$ will be of the order of $\sim 1/\sqrt{t_m \Delta f}$ where $t_m$ is the measurement time. For instance, for a single FFT with acquisition time $t_m$ the frequency resolution is $1/t_m$ and therefore there will be about $n=t_m \Delta f$ independent samples for the determination of $S_{ff}$. The same dependence on $t_m$ is obtained if FFT averaging is implemented. This simplified argument can be made more rigorous by means of Wiener filter theory~\cite{pontin}.

Now, let us assume that the detection is Heisenberg limited, i.e. $S_{ff} S_{xx}=\hbar^2$~\cite{braginsky} and that thermal noise is negligible. In this ideal case the force noise is completely determined by the measurement backaction, and therefore by the strength of the measurement. By using the Heisenberg condition to eliminate $S_{xx}$ from Eq.~(\ref{noisek}), we find:
\begin{equation}
  \Delta f = \frac{S_{ff}}{2 \pi m_s \hbar \omega_0} \label{df}
\end{equation}
We can connect this ideal measurement bandwidth with the phonon heating rate $\gamma$ due to $S_{ff}$, which is generally defined by the relation:
\begin{equation}
  \gamma = \frac{S_{ff}}{4 m_s \hbar \omega_0}. \label{hrate}
\end{equation}
By comparing Eqs.\,(\ref{df}) and (\ref{hrate}) we can conclude that $\Delta f =2 \gamma /\pi \simeq \gamma$, i.e. the effective measurement rate $\Delta f$ is set by the phonon heating rate $\gamma$.
In the nonideal case other force noise sources are present. We may define a factor $N$ such that $S_{ff}S_{xx}=  N^2 \hbar^2 $ and  performing a similar analysis we conclude that $\Delta f= 2 \gamma/\left( \pi N \right) \simeq \gamma/N$.


We turn now to the stroboscopic reheating strategy, which is described for instance in Ref.~\cite{goldwater,novotny}. Here one works in nonstationary conditions. In a first step the trapped particle is monitored with high sensitivity, and feedback-cooling is applied to prepare the system in a state as cold as possible with mean phonon number $\langle n_1 \rangle$. Subsequently, feedback is switched off and the oscillator will reheat due to the force noise $S_{ff}$. A possible advantage of this approach is that during the free reheating evolution one is allowed to switch off the detection, therefore avoiding completely measurement backaction. After a given evolution time $t$, the system energy is measured again and the heating rate $\gamma$ is inferred through the relation:
\begin{equation}
  \gamma= \frac{\langle n_2 \rangle - \langle n_1 \rangle}{t}
\end{equation}
where $\langle n_2 \rangle $ is the mean phonon after the free evolution.

How fast can a measurement of $\gamma$ and hence $S_{ff}$ be? Again, let us start with an ideal Heisenberg limited detector, in absence of other noise sources. Under these ideal conditions one may in principle perform feedback-cooling to the ground state, so that in the initial state $\langle n_{1} \rangle \simeq 0$ with uncertainty $\sigma_{n_1}=1/2$. After a free evolution time $t$ the energy is measured again giving $\langle n_2 \rangle = \gamma t$ with uncertainty $\sigma_{n_2}\simeq \sqrt{\langle n_2 \rangle}$. The relative uncertainty on the estimation of $\gamma$ will be:
\begin{equation}
  \frac{\sigma_\gamma}{\gamma} \simeq \frac{1}{\sqrt{\langle n_2 \rangle}} = \frac{1}{\sqrt{\gamma t}}.
\end{equation}
From a statistical point of view we can thus interpret $\gamma t$ as an effective number of independent samples for the experimental estimation of $\gamma$, so that $\gamma$ can be interpreted as an effective measurement rate. We have thus arrived essentially at the same expression of the continuous measurement case, apart of constants of order $1$, meaning that the time required to estimate $S_{ff}$ or equivalently $\gamma$ with a given accuracy is the same. In case of non ideal detection, we proceed as before by defining a factor $N$, and the initial state and the final state will be affected by a larger uncertainty. As in the continuous case, this leads to a reduction of the measurement rate to $\gamma/N$.

In conclusion, the two strategies appear to be roughly equivalent from a fundamental point of view. The choice of one instead of the other one will depend mostly on technical implementation aspects. Continuous strategies are in principle easier to implement, being based on a stationary state. However, measurements at very low coupling could be challenging for technical reasons and require high stability of the trap frequencies. Stroboscopic measurements do not suffer of the last problem but require the ability to deal with the transients associated to switching detection on and off, which can be extremely challenging in practice.

\subsection{Optical cavity}\label{sec_cav}

Here we consider an optical cavity exploited as a pure displacement sensor. Contrary to the tweezer approach, the cavity is sensitive mainly to one degree of freedom of the particle motion.  The requirements to achieve a meaningful measurement in the CSL context are quite stringent and will drive the design of the detection in a direction that is quite different from the typical opto-mechanical framework.

\begin{figure}[!ht]
\includegraphics[width=8.6cm]{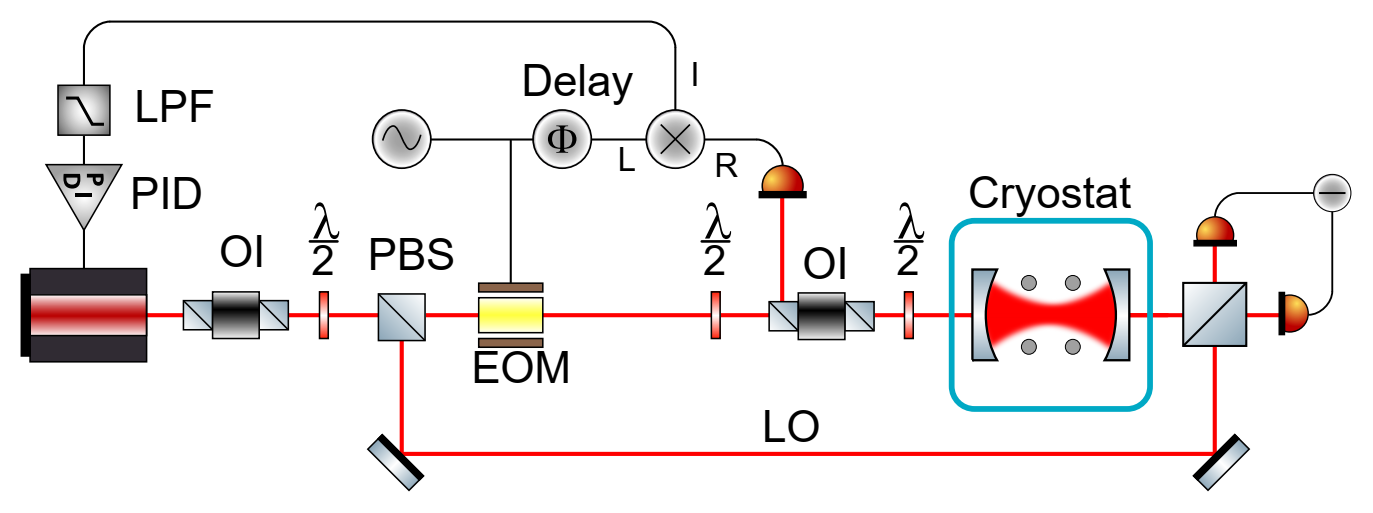}
\caption{Simplified scheme of the experiment. A Nd:Yag laser at $1064\,$nm is locked on the fundamental mode of an optical cavity by means of a standard Pound-Drever-Hall technique. A silica bead is held at the center of the cavity field by a linear Paul trap. Measurement of the particle dynamic is obtained by homodyning the transmitted cavity field.}
\label{scheme}
\end{figure}

We show in Fig.\,\ref{scheme} a schematic view of the experiment. A nano-particle is held at the center of an optical cavity by a Paul trap. The transmitted cavity field is then analyzed by homodyne detection. In general, the presence of a dielectric in an optical cavity causes the optical resonance to be downshifted in frequency. For a spherical Rayleigh particle this is simply given by~\cite{chang,barker} $\delta\omega=g_o\,cos[\phi]^2$,where $\phi=k x$ is the position of the particle in the cavity standing wave, with $k$ the wave number, and $g_o$ is the characteristic opto-mechanical coupling strength. Thus, the particle dynamics can be monitored by an optical phase sensitive detection. The highest sensitivity is obtained at $\phi=\pi/4$ (maximum of $\frac{\partial\delta\omega}{\partial x}$) where, to first order, the transduction is linear. At the same time, however, dipole forces provide a trapping potential with a trap frequency at the antinode ($\phi=0$) given by\,\cite{tania}:
\begin{equation}
  \omega_t=\frac{2 \hbar k^2 g_o}{m_s} n_c.
\end{equation}
\noindent Here, $n_c$ in the intra-cavity photon number. This potential is exploited in many applications~\cite{Ptrap,giacomo,kiesel} here, however, it represents an unwanted perturbation that can easily dominate over the Paul trap potential.

This very simple description already allows us to formulate two stringent requirements. First, assuming that the Paul trap and the cavity can be accurately aligned so that $\langle\phi\rangle=\pi/4$, the rms displacement needs to be small enough to keep the cavity transduction in the linear regime. This displacement has to include both the thermal secular motion and the driven micromotion.  Second, the optical potential needs to be negligible compared to the Paul trap potential. At the optimal position for the detection the optical potential exerts a force displacing the steady state position of the particle moving it away from the center of the Paul trap potential. As a consequence we need $\omega_t<\omega_0$. This requirement strongly limits the maximum intra-cavity power allowed, pointing to a low finesse cavity as the best choice.

Another critical aspect is the back-action introduced by the detection which will ultimately limit the force noise sensitivity. The main sources of back-action are radiation pressure shot noise and recoil heating\,\cite{pflanzer}, however, cavity dynamical effects must be accounted for since a non vanishing detuning will introduce optical spring and damping.

In the following we are going to show the expected sensitivity assuming an ideal homodyne detection, which is the best case scenario due to technical aspects. Laser frequency noise represents a major technical limitation. For a resonant optical drive, it can be considered as an additive noise limiting the sensitivity far above the shot noise\,\cite{pontin2}. In order to leave unaltered the particle signal in the homodyne, the cavity lock bandwidth has to be much smaller than $f_0$. This imposes an extremely demanding requirement in terms of displacement noise of the cavity mirrors, especially considering the cryogenic environment.

We consider an asymmetric low finesse ($\sim1000$) cavity under-coupled on the injection side.  The cavity length is $L=15\,$mm, has a waist of $62\,\mu$m and is coupled to a $200\,$nm radius silica nano-particle.  Assuming a $1064\,$nm laser driving the cavity, we have a single photon opto-mechanical coupling $g/2\pi=x_{zpf} k g_o/2\pi\simeq5$\,Hz. We consider optical input powers ranging from $0.1$ to $20\,\mu$W for which we can estimate the power absorbed by the particle. Assuming a pressure of $10^{-13}\,$mbar, which should be achievable in an ultra-cryogenic environment~\cite{drewsenP}, the particle equilibrium temperature is expected to be between $20$\,K to $70\,$K following a $W_{abs}^{1/5}$ power law since it is completely determined by black body radiation, as shown in section\,\ref{env}.

\begin{figure}[!ht]
\includegraphics[width=8.6cm]{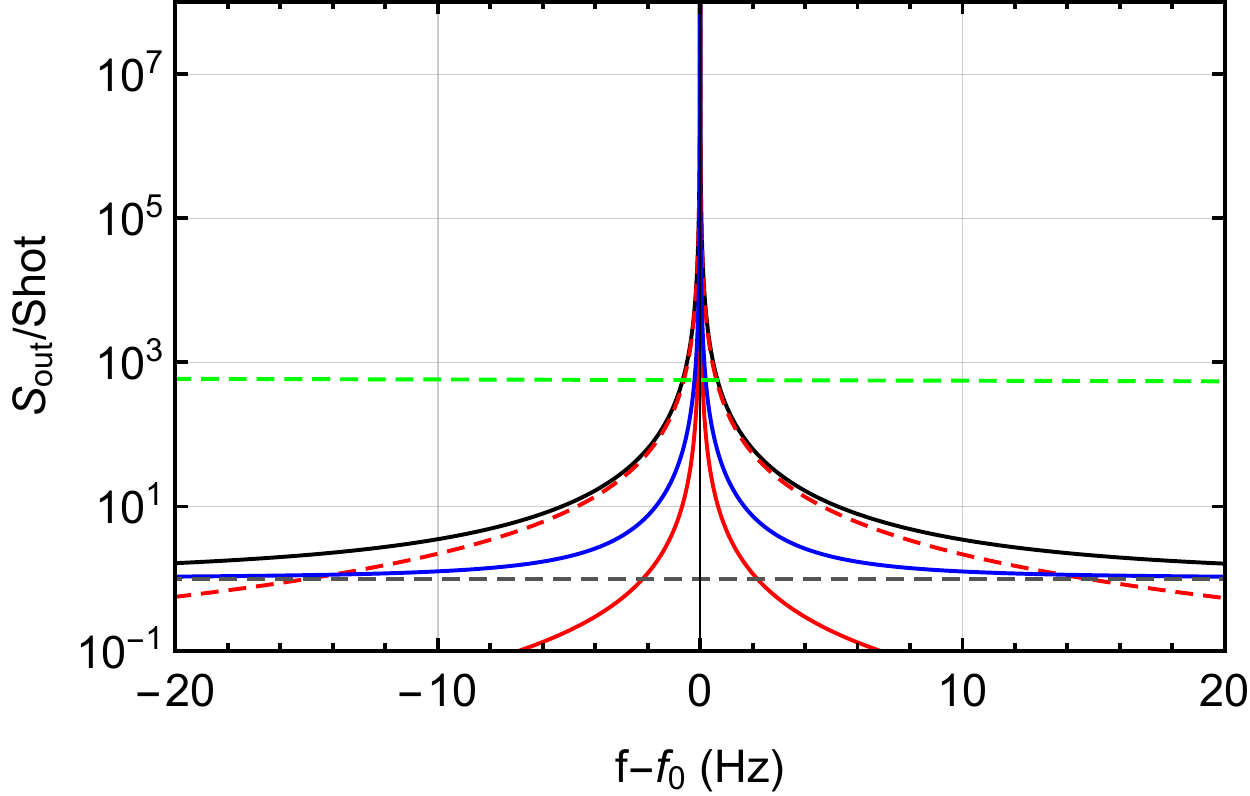}
\caption{Cavity output homodyne spectrum normalized to shot noise. Black is the total, blue the quantum noise, red thermal and red-dashed is photon recoil. Finally, the dashed green line represents the typical frequency noise that could hinder the particle detection.}
\label{cavspectrum}
\end{figure}

We show in Fig.\,\ref{cavspectrum} the expected PSD of the homodyne detection of the transmitted beam normalized to the shot noise for an input power of $10\,\mu$W along with all main contributions. Despite an effective bath temperature of $\sim20\,$K, the total PSD is dominated by photon recoil. This remains true for all input powers considered. Also shown in Fig.\,\ref{cavspectrum} is the estimated limit to sensitivity due to laser frequency noise. Typical spectra for high stability Nd:Yag lasers give $S_{\nu\nu}(f)\simeq  4 \times 10^{8}/f^2$\,\,Hz$^2/$Hz.

\begin{figure}[!ht]
\includegraphics[width=8.6cm]{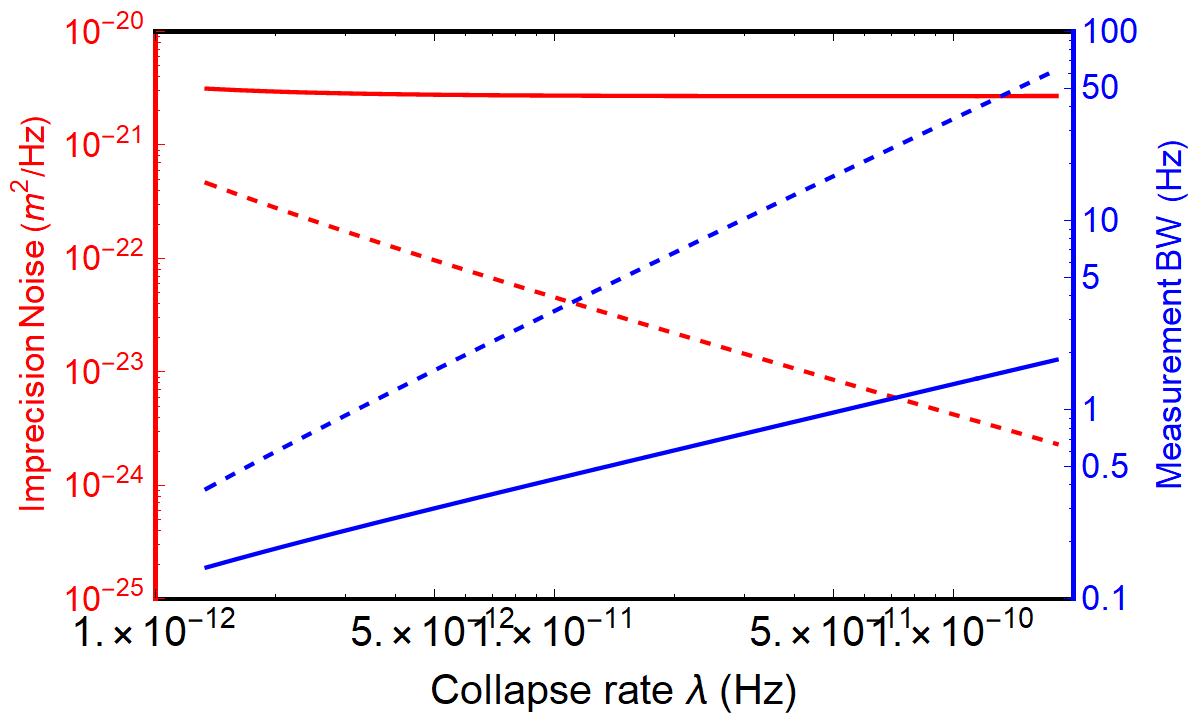}
\caption{Imprecision noise and measurement bandwidth as a function of the collapse rate $\lambda$ that would provide a signal to noise ratio of one given the estimated total force noise acting on the particle and assuming $r_c=10^{-7}\,$m. Continuous lines refer to a frequency noise limited detection while dashed line refer to a shot noise limited one. See main text for more details.}
\label{cavBW}
\end{figure}

In order to compare the sensitivity to CSL, we calculate the total force noise acting on the particle and express it in terms of collapse rate $\lambda$ assuming a characteristic length $r_C\,=\,10^{-7}$\,m. Therefore, the collapse rate that will give a signal to noise ratio of one, as is summarized in Fig.\,\ref{cavBW} where we plot the imprecision noise as a function of the collapse rate for both shot noise limited and frequency noise limited detection. For the former case, imprecision noise decreases for an increasing power as is typical for a shot noise limited detection. For the latter case, the imprecision noise is relatively constant. However, the CSL noise sensitivity is always reduced for an increasing power since the system is dominated by photon recoil.

Another important parameter is the measurement bandwidth since it directly impacts measurement time and as a consequence the requirements on stability on all other parameters. This is shown in Fig.\,\ref{cavBW} as well. For a frequency noise limited detection this turns out to be $\sim150\,$mHz for the smaller power and up to $\sim2\,$Hz for the highest. For shot noise limited readout the situation would be much more favorable with a bandwidth varying from $\sim400\,$mHz to $\sim60\,$Hz. Of course, achieving this seems rather challenging especially considering the cryogenic environment. Indeed, for the configuration considered, the shot noise level would correspond to a relative displacement noise of the cavity mirrors of the order of $\sim10^{-33}\,$m$^2/$Hz.

So far we have assumed a rather optimal scenario, here we discuss some criticalities. First, as stated before, the rms displacement of the particle needs to be sufficiently small to remain in the linear transduction region of the cavity. That is $\Delta x = (k_B  T_{eff}  /m \omega_0^2)^{1/2}\leq\lambda/16$. For the parameters considered here we have $\Delta x\simeq0.4\lambda$ meaning that additional active feedback is required in order to reduce the particle fluctuations. To meet this requirement the mechanical quality factor needs to be reduced by a factor $100$. This seems reasonable considering that its value is expected to be $\sim10^9$ at $P_g=10^{-10}\,$mbar and increases inversely proportional to pressure.

The optical spring near resonance has a linear dependence on detuning we can impose a shift in frequency equal to the mechanical linewidth $\Gamma_{tot}$ from which we can get a limit on the detuning for given cavity parameters. That is

\begin{equation}
  \Delta\leq\frac{c }{2 P_{in}}\frac{m_s w_0 \kappa^4}{2g_o^2 k \kappa_{in}} \Gamma_{tot}
\end{equation}

\noindent where $k$ is the wave number, $\kappa$ is the cavity half linewidth and $\kappa_{in}$ the contribution to it due to the input port. For our parameters and an input power of $10\,\mu$W this is roughly $\Delta\leq10^{-7} \kappa\simeq 0.4$\,Hz which is an extremely demanding requirement. Just to give a reference the maximum shift is $\sim1$\,Hz. Since we are in the deeply bad cavity regime the same kind of limit imposed using the optical damping is much less stringent. The situation becomes more relaxed if we assume some kind of active feedback increasing the mechanical linewidth. However, this increase needs to be by a factor $\sim1000$ at least. At the same time, any blue detuning will give rise to dynamical instability. Thus a small but finite detuning seems rather necessary.

Some final considerations are required concerning the optical power considered. The plots in Fig.\,\ref{cavBW} summarizing the performance in terms of sensitivity to CSL noise consider input powers ranging from $0.1$ to $20\,\mu$W. The upper bound to the input power is due to the optical potential as previously discussed. The lower bound is somewhat more free but will ultimately be set by technical considerations concerning cavity locking. Indeed, a simple locking scheme as depicted in Fig.\,\ref{scheme} is quite challenging to implement with the minimum power considered. However, more complicated locking schemes could be implemented. For example the cavity could be locked to higher order cavity mode (i.e. a TEM$_{01}$) which has a node along the cavity axis and is thus not coupled to the particle to first order, while a second laser is offset locked to be resonant to the cavity fundamental mode. A detailed study of the optimal optical setup is not the main focus of this paper and will be the topic of future research.

\subsection{Optical tweezer}\label{sec_tweezer}

\begin{figure}[t]
\centering \includegraphics[width=1\columnwidth]{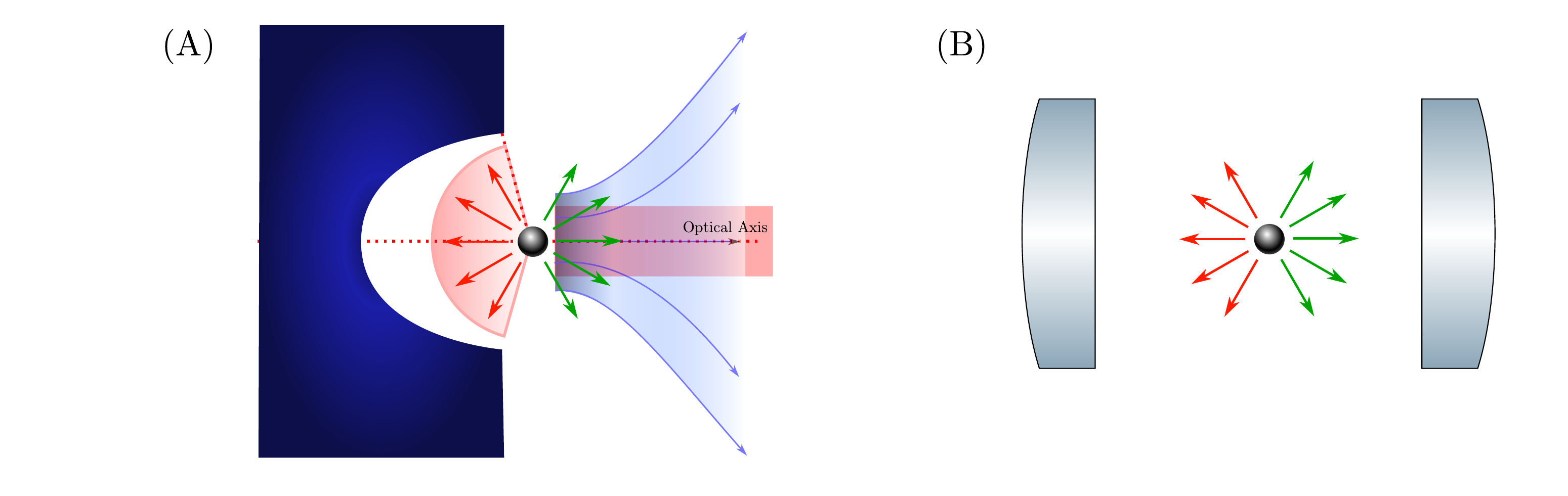} \caption{\label{fig:setup} \textbf{(a)} Detection using a paraboloidal mirror,
and \textbf{(b)} detection using an objective lens setup.}
\end{figure}

We consider two different tweezer-based detection setups depicted in Fig.~\ref{fig:setup}. Specifically, we consider a nanoparticle in a Paul trap with trap frequency $\omega_{\text{0}}$ which is monitored either continuously or stroboscopically using a tweezer. The trapped nanoparticle scatters light which is collected using optical elements and then directed towards the detector; similarly as in the cavity case we again consider the situation where the scattered light interferes with a local oscillator in a homodyne scheme~\cite{rashid2017wigner}.
For concreteness, we consider a laser wavelength $\lambda=1550$\,nm
and assume a spherical silica ($\text{Si}\text{O}_{2}$) nanoparticle of radius $R=200$\,nm.

The Paul trap potential, $U_{\text{Paul}}$, can be significantly perturbed by the optical potential, $U_{\text{optical}},$ as well as by the effective potential $U_{\text{scatt}}\propto F_{\text{scatt}}z$ generated by the non-conservative scattering force $F_{\text{scatt}}$ oriented in the propagation direction $z$ of the laser beam~\cite{timberlake2019static}. The presence of the optical potential $U_{\text{opt}}\propto\omega_{\text{optical}}^{2}\propto P$ results in a change of the trap frequency, i.e. $\omega_{\text{0}}^{2}+\omega_{\text{opt}}^{2}$, where $\omega_{\text{opt}}$ and $P$ denote the optical frequency and laser power, respectively. On the other hand, the effective scattering potential $U_{\text{scatt}}$ displaces the origin of the combined Paul and optical trap; if the displacement is too large the nanoparticle can leave the harmonic region of the trap, possibly even resulting in particle loss. For the considered nanoparticle of size $R=200$\,nm, noting that $U_{\text{optical}}\propto R^{3}$ and $U_{\text{scatt}}\propto R^{6}$, we find the strongest constraint coming from the requirement $\vert\partial_{z}U_{\text{scatt}} \vert\apprle\vert\partial_{z}U_{\text{Paul}}\vert$; specifically, assuming a Paul trap frequency $\omega_{0}=2\pi\times1$\,kHz we are limited to powers of $P\apprle1$\,nW. To avoid the problem related to the scattering potential $U_{\text{scatt}}$ one can however consider two counter-propagating beams which could be implemented using the lens setup shown in Fig.~\ref{fig:setup}(b); requiring then $U_{\text{optical}}<U_{\text{Paul}}$ we then find a much less restrictive condition on the laser power, i.e. $P\apprle1$\,mW.

\begin{figure}[!ht]
\includegraphics[width=8.6cm]{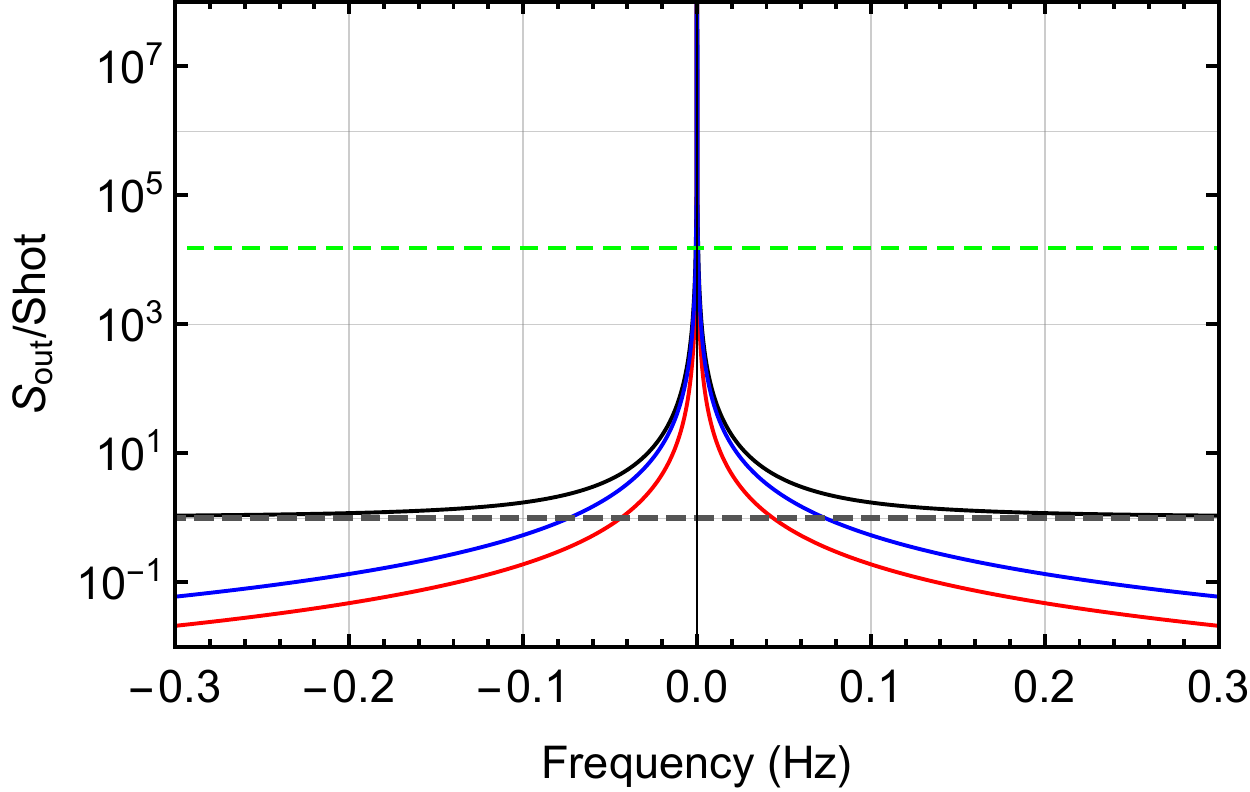}
\caption{Optical tweezer homodyne spectrum normalized to shot noise. Black is the total noise, blue the quantum noise (photon recoil and backaction), red thermal.
Finally, the dashed green line represents the noise floor estimated using NEP to quantify the detector's dark noise which is expected to significantly hinder the detection.}
\label{fig8}
\end{figure}

Ideally we would like to set the laser power $P$ to the value which minimizes the sum of the backaction and imprecision noise~\cite{braginsky}. However, the optimal value of $P$ is relatively low and sources of dark noise present in physical detectors must be taken into account. From the laser power $P$ and the particle position $z$ one can then estimate the photocurrent impinging on the detector; we quantify it by the total efficiency $\eta$, and denote the detected power by $P_{\text{det}}$.
In particular, adopting a simple semi-classical calculation we can find the conversion factor between the detected power and the particle position~\cite{rashid2017wigner}. We compare $P_{\text{det}}$ to the detector's dark noise, which can be estimated by considering the noise equivalent power (NEP) of the detector~\cite{vovrosh2017parametric}. The NEP noise floor, although not an intrinsic limitation of the
tweezer setup, is expected to be the dominant contribution to the noise floor (see Fig.~\ref{fig8}).

The fundamental noise floor in a tweezer setup, similarly as in the cavity setup, is given by the shot noise. The nanoparticle motion is affected by noise from gas collisions, photon-recoil and backaction. In Fig.\,\ref{fig8}, we show the expected homodyne spectrum for the tweezer case. We are assuming a pressure of $10^{-13}$\,mbar, a numerical aperture $NA=0.6$ limited by reasonable trap geometries and an input power of $500$\,nW at a wavelength of $1550$\,nm. With this power we estimate a particle equilibrium temperature of $T\simeq60$\,K, quite similar to the cavity case shown in Fig.\,\ref{cavspectrum}. Already at such low power the particle dynamics is dominated by quantum noise.

\begin{figure}[!ht]
\includegraphics[width=8.6cm]{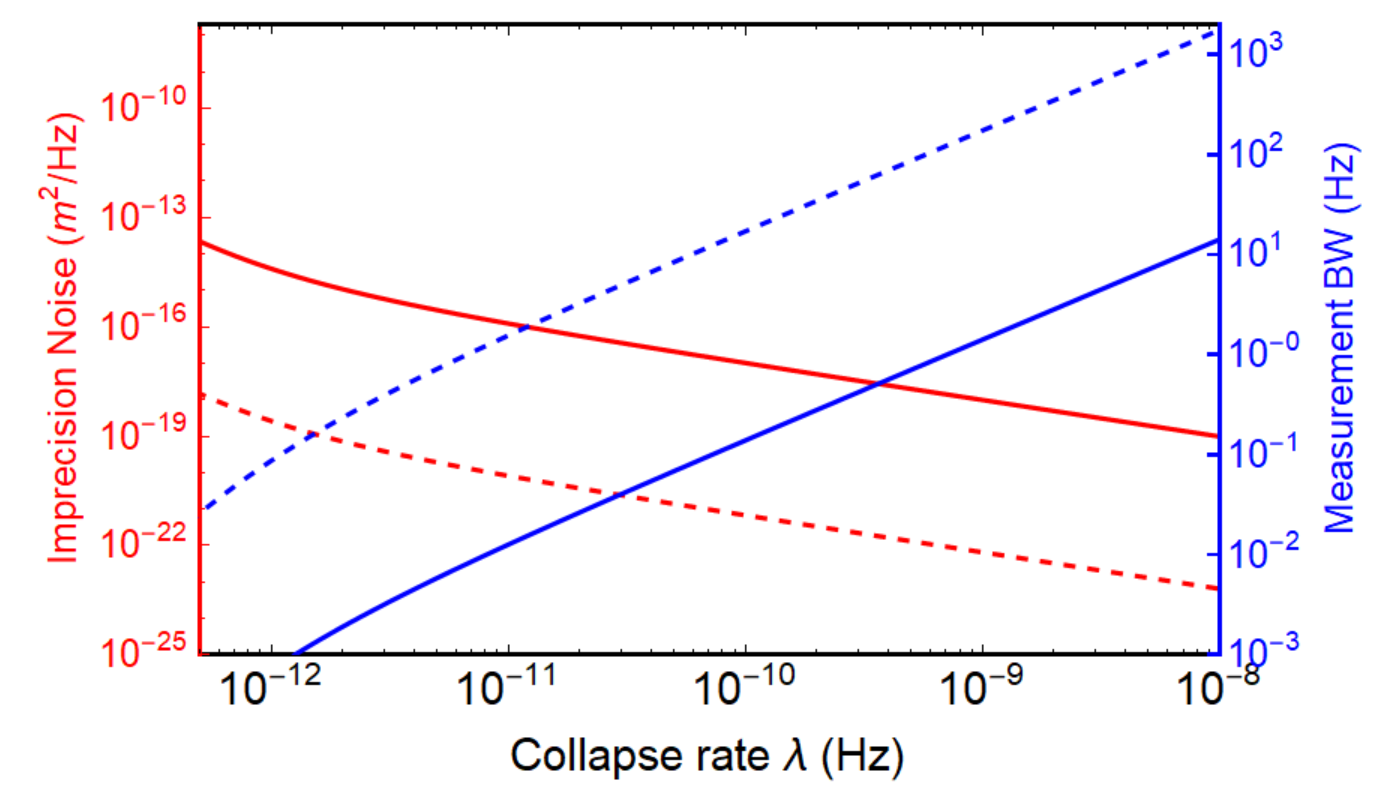}
\caption{Imprecision noise and measurement bandwidth as a function of the collapse rate that would provide a signal to noise ratio of one given the estimated total force noise acting on the particle and assuming $r_{C}=10^{-7}$\,m. Continuous lines refer to a NEP noise limited detection while dashed line refer to a shot noise limited one.}
\label{fig9}
\end{figure}

To summarize the performance and allow an easy comparison, we show, in Fig.\,\ref{fig9}, the imprecision noise and measurement bandwidth as a function of detectable $\lambda$, in analogy to Fig.\,\ref{cavBW}. The main advantage of the tweezer lies in the much wider range in acceptable power. Contrary to the cavity case, the maximum sensitivity is obtained in the position of maximum intensity of the optical field. This, allows to fully exploit the cancelling forces due to the counter-propagating beams. However, the resulting displacement sensitivity is lower with a direct impact to the attainable measurement bandwidth.

The most critical aspect of the tweezer approach lies in the requirements on the positioning accuracy of the two objective lenses. In order for the suppression of the scattering force to be effective, the position and size (i.e. non idealities) of the waist of the two lenses needs to be close to a fraction of the nominal waist which is of the order of $\sim2\,\mu$m (i.e. half the radius of the core of a telecom singlemode fiber). While this is standard practice at room temperature, maintaining the alignment in an ultra-cryogenic environment is a completely different endeavour.


\subsection{Electrical readout with a SQUID}\label{sec_squid}

Instead of detecting the oscillating particle using an optomechanical setup, it has been proposed to use a direct electrical detection~\cite{goldwatermillen}. To this end, a pair of electrodes, for instance two endcaps, are used to detect the motion of the particle along the axis orthogonal to them. The electrical signal induced in the electrodes could be eventually readout by a SQUID current sensor. SQUIDs are the best electrical amplifiers available in a wide range of frequencies, and are natural choice when working at low temperature and low frequency.

Unfortunately the electromechanical coupling of the above scheme is very weak for a current-based detection at low frequency. As shown in\,\cite{goldwatermillen,wineland}, the motion $x$ of the particle with charge $q$ will induce a current in the circuit connected to the electrodes:
\begin{equation}
  I=\frac{q}{d} \dot{x}=\frac{\omega q}{d} x;  \label{transduction}
\end{equation}
where $d$ the effective gap between the electrodes and $\omega$ is the oscillation frequency. Furthermore, it was shown that the mechanical oscillator, as seen from the electrical circuit, is dynamically equivalent to a RLC oscillator with effective inductance $L_m=m_s/\beta^2$ and capacitance $C_m=1/\omega_0^2 L_m$, where $\beta=q/d$.

For a reasonable set of parameters envisaged for the experiment, $m_s=6\times 10^{-17}$\,kg (standard $200$\,nm radius SiO$_2$ particle), $q=10^3\,e$, $d=300$\,$\mu$m, $\omega_0/2 \pi=1$\,kHz one obtains effective LC inductance and capacitance $L_m=2 \times 10^8$\,H and $C_m=10^{-16}$\,F. For comparison a typical SQUID sensor has an input inductance $L_i \simeq 10^{-6}$\,H, which means an impedance mismatch by $14$ orders of magnitude! In fact, with a typical input current noise $S_i \simeq 0.1$\,pA/$\sqrt{\mathrm{Hz}}$, a direct coupling of the electrodes to a SQUID would lead to a really poor displacement resolution $S_x \simeq 30$\,$\mu$m/$\sqrt{\mathrm{Hz}}$.

One of the reasons for such huge mismatch is that, according to Eq.\,(\ref{transduction}), the displacement to current transfer function is proportional to $\omega$, which makes this scheme very suboptimal at low frequency. This situation is analogue to Faraday detection of magnetic fields.

To overcome this problem, one can explore different strategies. The first is to reduce impedance mismatch, the second is to resort to other types of amplifier, such as standard FET transimpedance amplifiers or single electron transistors.

\subsubsection{SQUID with untuned transformer}  \label{untuned}
\begin{figure}[!ht]
\includegraphics[width=8.6cm]{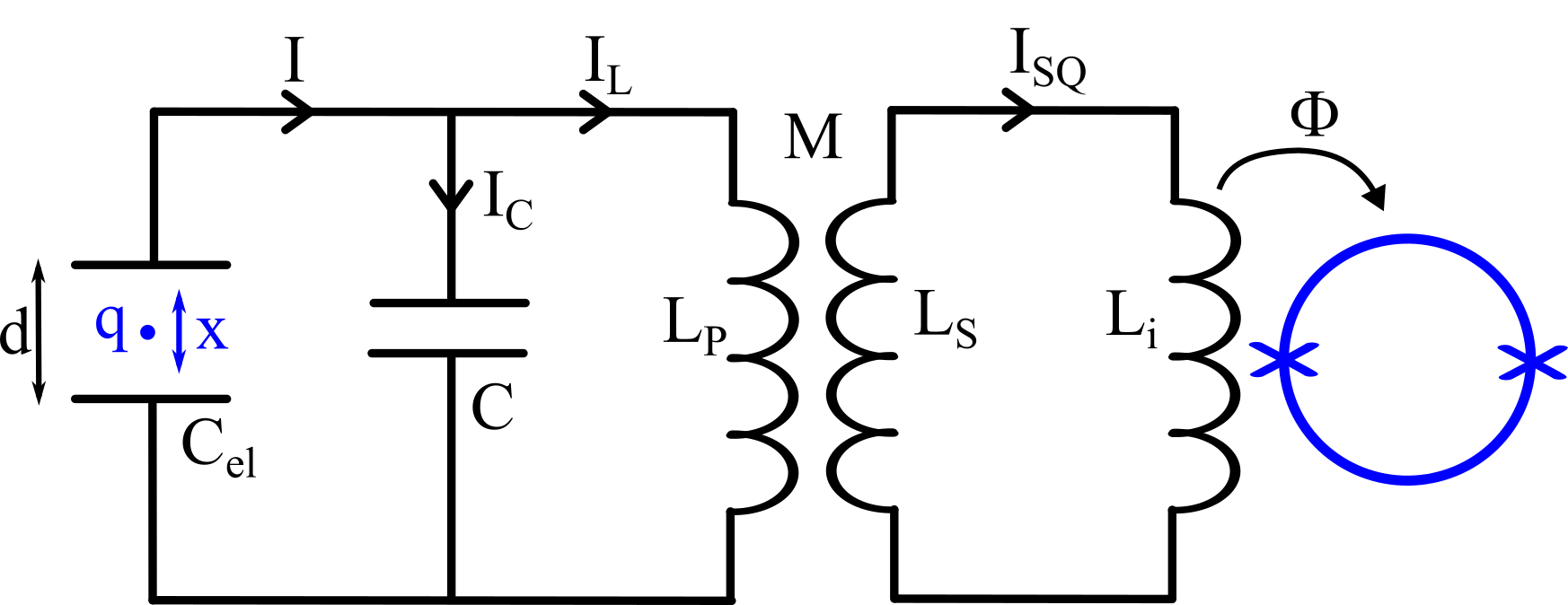}
\caption{Scheme of a SQUID-based detection of the levitated particle. The motion of the charge $q$ induces a current $I$ in the electrodes of capacitance $C_{el}$. The current is enhanced by a low-loss superconducting transformer with mutual inductance $M$ and high inductance ratio ($L_p \gg L_s$) before being measured by a SQUID of input inductance $L_i$ ($L_i \simeq L_s$). A further way to boost the current and improve impedance matching is to add a capacitor $C$ ($C \gg C_{el}$) in order to tune the LC resonance to the mechanical resonance of the trapped particle. }
  \label{figSQUID}
\end{figure}

The simplest way to reduce impedance mismatch to a SQUID is to interpose a superconducting transformer between the electrodes and the SQUID, as shown in Fig.\,\ref{figSQUID}. This increases the effective input inductance to a value of the order of the inductance $L_p$ of the primary coil. Low-loss superconducting  transformers with inductance up to $10$\,H, $Q \simeq 10^6$ and volume of around $1$ liter have been used to match a SQUID to the capacitive transducer of resonant bar gravitational wave detectors~\cite{falferi}. Larger values are in principle achievable but are hardly compatible with standard cryostats.

Let us consider as a maximum realistic choice $L_p=10$\,H, and assume the transformer geometrical coupling $k=M^2/(L_p L_s)$ to be close to the maximum allowed value of 1. For optimal matching the secondary coil inductance should be $L_s \simeq L_i$, where $L_i$ is the SQUID input inductance.

The SQUID can be modeled as a current amplifier with an imprecision current noise, with spectral density $S_{II}$, and a conjugate backaction voltage noise with spectral density $S_{VV}$. Neglecting crosscorrelations, a rough approximation for the spectral densities is $S_{II} \simeq 2 \hbar/L_i N_{\hbar}$ and $S_{VV} \simeq 2 \hbar \omega^2 L_i N_{\hbar}$. This way a single noise parameter $N_{\hbar}$ is singled out. The best available SQUIDs have shown effective noise $N_{\hbar} \simeq 10$ which is not far from the Heisenberg limit $N_{\hbar} = 1$.

Besides SQUID noise, one has to consider the Nyquist electrical thermal noise of the transformer. At kHz frequency the best reported electrical quality factor is $Q \simeq 10^6$~\cite{falferi2}. This implies a voltage noise referred to the primary coil $S_{VVel}=4 k_B T_g \omega_0 L_p/Q$. Taking into account the transformer coupling, Nyquist noise is equivalent to a SQUID backaction noise with equivalent noise number $N'_{\hbar} \simeq 2 k_B T_g/(\hbar \omega_0 Q)$. It happens that for $T_g=300$\,mK and $Q \simeq 10^6$, electrical noise is slightly larger than SQUID backaction noise at $1$\,kHz. Thermal noise is dominant for $f_0 < 1$\,kHz, and becomes negligible for $f_0 \gg 1$\,kHz.

\begin{figure}[!ht]
\includegraphics[width=8.6cm]{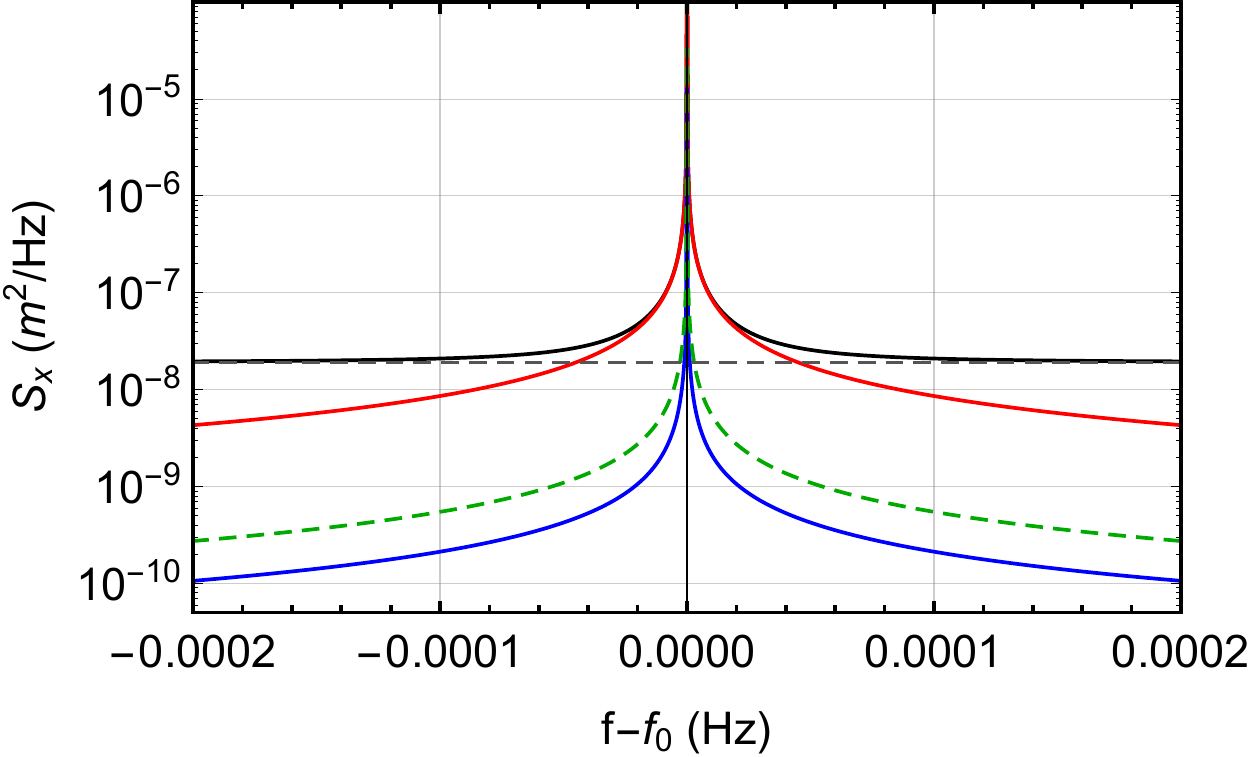}
\caption{Spectrum of a levitated particle measured by a SQUID through a superconducting transformer. The particle is a silica nanosphere with $R=200$\,nm, charge $q=10^3\,e$, the resonance frequency $f_0=1$\,kHz, the residual gas is helium at $T_g=300$\,mK and $P_g=10^{-12}$\,mbar. The SQUID noise is $N_{\hbar} = 10$ and the transformer has $L_p=10$\,H, coupling $k=0.8$ and $Q=1\times10^6$. The various noise components are: total noise (solid black), gas collision noise (solid red), electrical noise (dashed green), SQUID backaction (solid blue) and SQUID imprecision (dashed gray). The total force noise around resonance corresponds to a CSL collapse rate $\lambda \simeq 6 \times 10^{-14}$\,Hz.}
  \label{figSQ1}
\end{figure}

Having defined the parameters, we can now simulate the spectrum of a levitated nanoparticle coupled to a SQUID through a transformer. Fig.\,\ref{figSQ1} shows a spectrum based on the representative set of parameters discussed above. The spectrum shows that the coupling would be so low that both Nyquist noise and backaction noise are much smaller than the gas collision noise, even at a pressure $P_g=10^{-12}$\,mbar. Assuming the particle to be thermalized, as there is no direct dissipation of energy, the force noise on resonance would allow to test the CSL model down to $\lambda<10^{-13}$\,Hz. The drawback is that because of low coupling, the bandwidth would be extremely narrow, $\Delta f < 1$\,mHz.

\begin{figure}[!ht]
	\includegraphics[width=8.6cm]{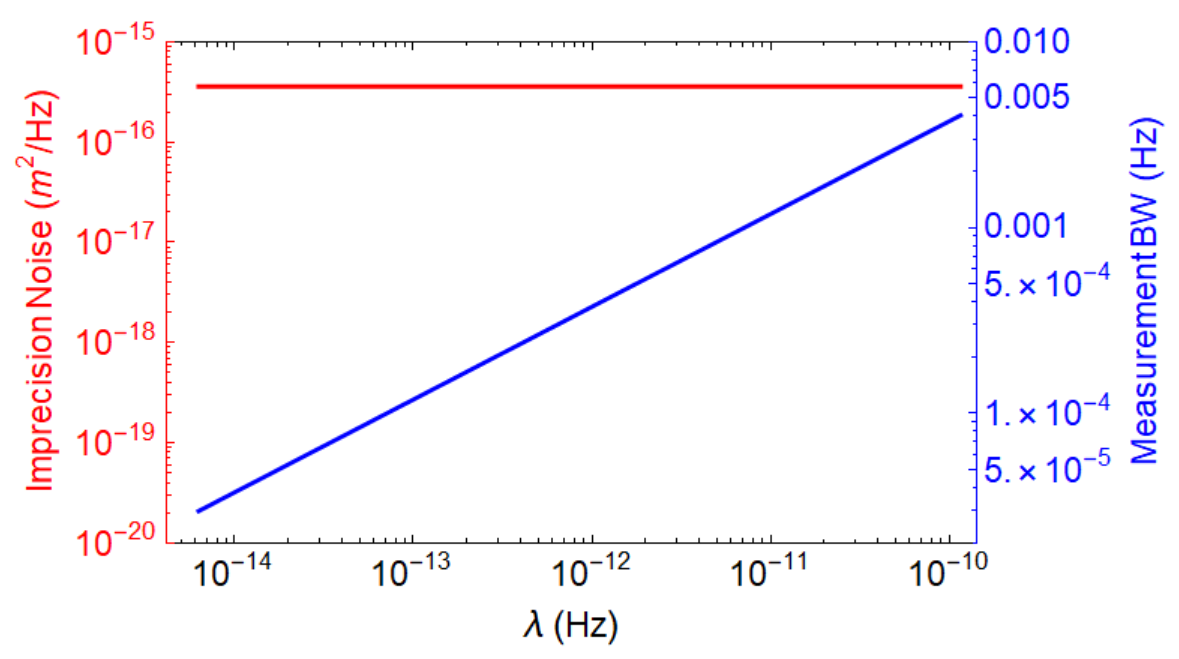}
\caption{Displacement noise and measurement bandwidth, expressed as a function of the minimum measurable collapse rate $\lambda$, evaluated at the standard value $r_C=10^{-7}$\,m, for the untuned SQUID readout. The parameters are the same as in Fig.~\ref{figSQ1}, except of the gas pressure which is variable, in order to vary the force noise and thus the detectable $\lambda$.}
\label{SQUIDBW}
\end{figure}

For the sake of comparison with the optical cavity and tweezer approaches, we construct a bandwidth vs detectable $\lambda$ in analogy to Fig.~\ref{cavBW}. In contrast with the optical case, the back action noise here is negligible and the position noise is fixed by the circuital parameters, so that there is no way to tune the measurement coupling (as changing power in the optical case). We vary the thermal force noise by varying the gas pressure $P$ , and calculate the corresponding minimum detectable $\lambda$ and bandwidth. In this way we obtain the plot in Fig.~\ref{SQUIDBW}.

Clearly, the striking feature of a SQUID-based detection of a nanoparticle in a Paul trap is the extremely narrow bandwidth. In principle such an experiment would not be unfeasible, as it would be possible to acquire tens of independent points in one day. However, this would require the trap frequency to be extremely stable within better than 1 mHz over such a time scale. This is not trivial for an actively controlled trap. In addition, note that the experimental parameters are already quite tight. Lower charge, higher electrode distance or lower primary coil inductance will all have the effect of further reducing the bandwidth.

Despite the very narrow bandwidth, a SQUID-based detection has the very attractive feature of being effective with any type, material or size of the charged particle. It would then be possible to levitated much heavier particles, for instance made of gold or other heavy elements such as platinum or osmium, which would strongly increase the coupling to CSL. For instance, with a osmium particle with $R=400$\,nm and the same other parameters of Fig.\,\ref{figSQ1}, one would reach a force noise of $\lambda \simeq 3 \times 10^{-16}$\,Hz, close to the GRW limit. Unfortunately, in this case the measurement bandwidth would reach the extremely small value of $\Delta f \simeq 2$\,$\mu$Hz.

It is fair to say that, besides the very long measurement time, we have completely ignored here the voltage noise in the trap bias line. This is actually expected to become a big issue especially for large charge. To make this contribution negligible in the configurations discussed above, it would be necessary to suppress voltage noise in the bias line to $S_{VV}< 10$\,pV/$\sqrt{\mathrm{Hz}}$. Furthermore, we have to consider another nontrivial technical issue. The SQUID electronics must be able to handle the signal due to crosstalk from the ac bias line, which will likely be huge. While this crosstalk can be in principle suppressed by proper cancellation schemes, this may be not so easy to do in practice.

\subsubsection{SQUID with transformer and LC}

To further improve the impedance matching, one can tune a LC resonance to the resonance of the trapped particle. Note that a similar strategy at comparable frequency has been pursued in the readout of resonant bar gravitational wave detectors\,\cite{falferi}. Besides the increased technical difficulty, it has been shown that this tuned LC system is really advantageous in terms of bandwidth only if the electrical quality factor is comparable or better than the mechanical quality factor. While this condition was met in the case of Ref.\,\cite{falferi}, it is definitely not valid in our case, as the largest electrical quality factors at kHz are of the order of $10^6$, while the mechanical quality factor in the configurations here considered is expected to be of the order of $10^{11}$.

\begin{figure}[!ht]
\includegraphics[width=8.6cm]{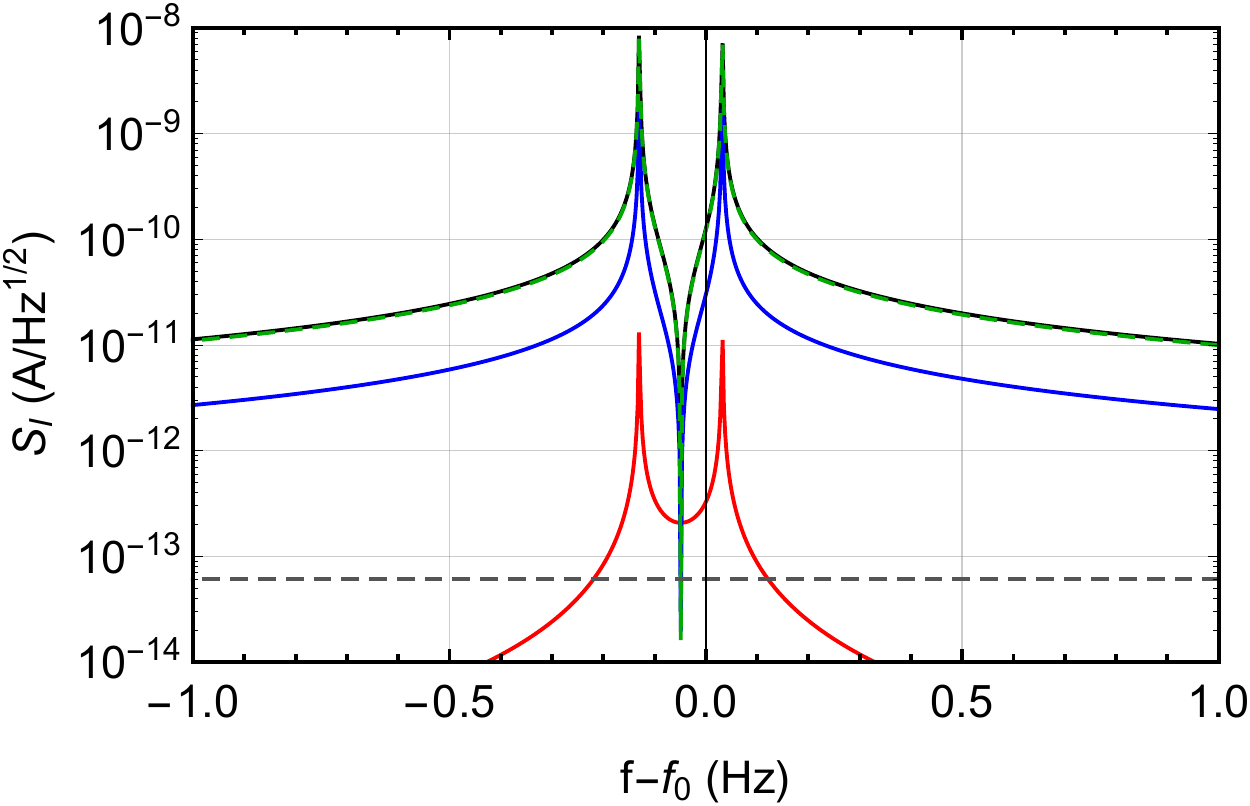}
\caption{Spectrum of a levitated particle measured by a SQUID through a superconducting transformer with tuned LC resonance. Parameters are exactly the same as in Fig.\,\ref{figSQ1}, except that a capacitor $C$ is added in parallel to the primary inductor to tune the electrical and mechanical frequencies. The spectrum shown here is expressed as current at the input of the SQUID. The color and line type coding for the different contributions is the same as in Fig.\,\ref{figSQ1}. Note that the electrical noise (dashed green) is hardly visible as it almost coincides with the total noise (solid black) everywhere except in the narrow antiresonance between the two peaks.}
\label{figSQ2}
\end{figure}

\begin{figure}[!ht]
\includegraphics[width=8.6cm]{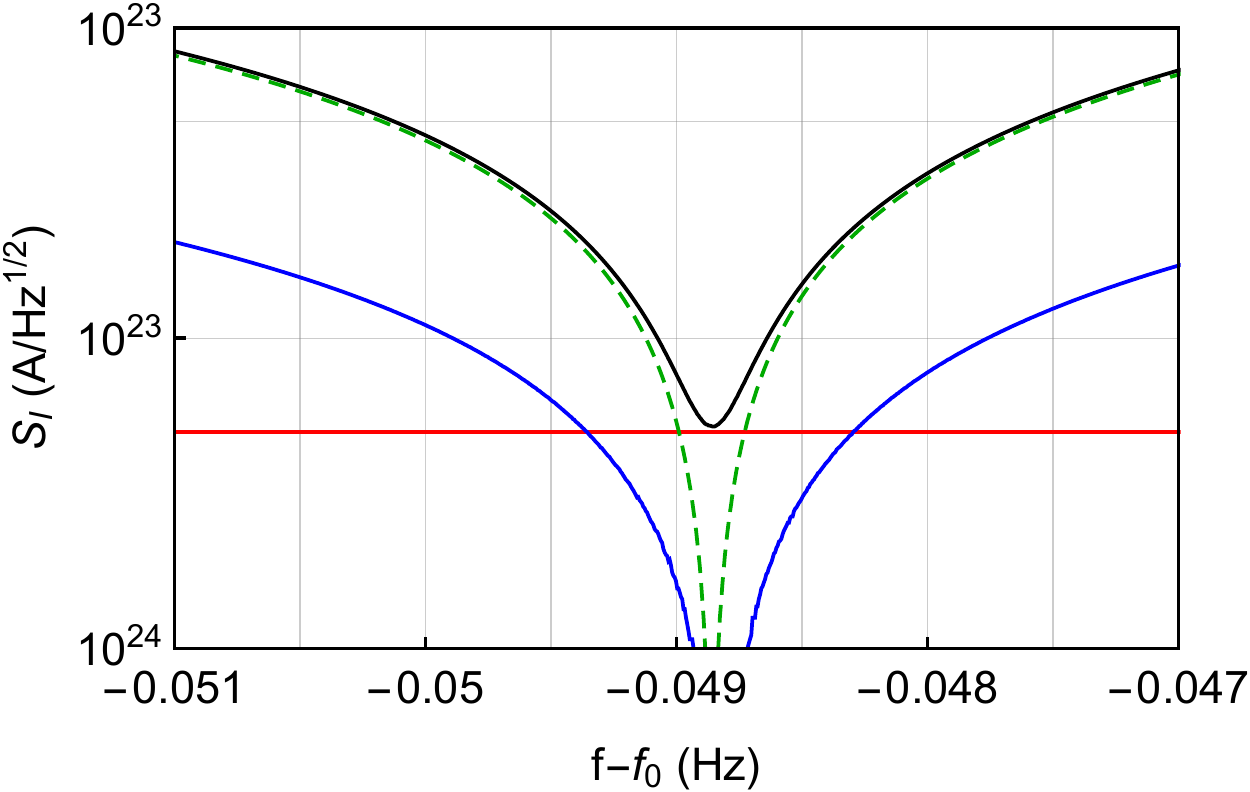}
\caption{Spectrum in the same situation of Fig.\,\ref{figSQ2}, but now expressed as force noise on the levitated particle, and zoomed in a narrow range around the deep antiresonance visible in Fig.\,\ref{figSQ2}.}
\label{figSQ3}
\end{figure}

To assess whether this intuitive argument is correct we have performed simulations of a circuit with the same parameters as in the transformer setup, but now the capacitor $C$ shown in Fig.\,\ref{figSQUID} in parallel to the primary inductor is used to tune the LC frequency of the circuit to the mechanical frequency.

The simulations is performed by explicitly writing the Kirchoff equations and the coupled mechanical equation, using as independent variable the mechanical displacement $x$, the currents across capacitor and inductor $I_C$ and $I_L$ and the current in the SQUID $I_{SQ}$ and including all relevant noise terms. The total output noise is then divided by the force to SQUID current transfer function to compute the effective total force noise.

First, we have checked that for large detuning between electrical and mechanical frequency, we recover the results of the untuned transformer circuit. Then, we have tuned the two frequencies and obtained the interesting behaviour shown in Figs.\,\ref{figSQ2} and\,\ref{figSQ3}. The two modes hybridize forming two coupled normal modes. The splitting depends on the electromechanical coupling, but even for large charge $q=10^3 e$ it is smaller than $1$\,Hz. In this situation, energy flows back and forth between electrical and mechanical resonators at a rate given by the frequency splitting. However, as the electrical quality factor (here $1 \times 10^6$) is much larger than the mechanical one (here $Q \simeq 3 \times 10^{11}$ for helium gas at $P_g=10^{-12}$\,mbar), the noise power is dominated by the former, largely degrading the force noise over almost the whole bandwidth.

Remarkably, there is a very narrow region between the two peaks where the electrical noise features a narrow antiresonance.
This point corresponds to the uncoupled mechanical resonant frequency. The antiresonance region is zoomed in Fig.\,\ref{figSQ3}, where the spectrum is expressed as force noise on the levitated particle. In a very narrow bandwidth of around $0.2$\,mHz the electrical noise drops below the very small gas-collision mechanical noise.

Within this bandwidth the force noise spectral density is again dominated by the gas, so it is the same as in the untuned case of Fig.\,\ref{figSQ1}. Remarkably, the effective bandwidth, despite the very different configuration, is substantially the same. This means that in terms of measurement speed the tuned LC shows no substantial advantage with respect to the untuned case. The situation would slightly improve if the electrical noise were improved by, say, two orders of magnitude. In this case the bandwidth would be limited by SQUID back action noise.

Finally, note that, similarly to the untuned case, see Fig.~\ref{SQUIDBW}, the bandwidth would be increased if the pressure were increased to values higher than $10^{-12}$\,mbar. This would imply larger force noise, i.e. worse detectable $\lambda$.

\subsubsection{Other types of amplifier}

One may also consider to use ultralow noise semiconductor amplifiers instead of SQUIDs. Transimpedance FET amplifiers feature an input current noise that can be as low as fA/$\sqrt{\mathrm{Hz}}$, thus even better than conventional SQUIDs. Unfortunately, in our experiment we need to consider the backaction noise as well. For a good transimpedance FET amplifier it is usually in the order of $1$\,nV/$\sqrt{\mathrm{Hz}}$. This is more than 6 orders of magnitude higher than a SQUID at $1$\,kHz. In fact, if we define the "noise number"  of a FET amplifier in analogy to a SQUID we find $N_{\hbar} \simeq 10^6$. This clearly makes this solution not viable.

A better case is constituted by single electron transistors. These are intrinsic charge sensors, in contrast with SQUIDs that are current sensors, so that the transfer function in a Paul trap setup (displacement to charge) would be flat. In particular, Superconducting Single Electron Transistors (SSETs) can be considered as a dual charge-based version of SQUIDs. They have demonstrated near quantum limited performance, with charge noise down to $10^{-6}$\,$e/\sqrt{\mathrm{Hz}}$ over an input impedance of the order of $1$\,fF, which would nicely match the apparent mechanical impedance of the trapped particle. Theoretically, this seems to be the optimal solution for the Paul trap readout.

Unfortunately, the development of SSETs has been much more limited with respect to SQUIDs. SSETs are not available at commercial or semicommercial level, and have been so far implemented only in very limited range of applications. In general they are much less robust and more sensitive to spurious effects, and it is questionable if there is any chance to operate a SSET in an application requiring huge dynamic range, such as in the readout of a trapped particle.

\section{Discussion}

In this section we will compare the different approaches quantitatively. We have to remark that all three experimental solutions require a substantial progress beyond current performance, and the solution of a lot of technical issues. In the following we optimistically assume a best scenario for all three cases, which leads to the bandwidth vs detectable $\lambda$ plots in the previous section.

%
%


By comparing Fig.~\ref{cavBW}, Fig.~\ref{fig9} and Fig.\,\ref{SQUIDBW} we draw the following conclusions:

\begin{enumerate}
	\item Even if detecting exceedingly low forces due to CSL requires extremely low power, compared to standard optomechanical experiments, optical readout allows a much more sensitive position measurement compared to SQUIDs, which translates into a larger bandwidth, i.e. speed of measurement. For a continuous monitoring, this becomes a crucial aspect if the trap features long term instabilities or drifts.
	\item The cavity performs better than the tweezer especially for very low values of lambda, down to $10^{-12}$\,Hz, however presents more technical limitations and less flexibility. The tweezer approach is more flexible and can be used, for instance, on a wider range of power. In both situations, technical noise is hardly avoidable, in particular laser frequency noise in the cavity and detector dark noise in the tweezer. However in the ultralow power limit the cavity performs better.	
	\item A SQUID approach is clearly unsuitable under the bandwidth criterium. However it would present some advantages: first, the power dissipated in the nanoparticle by the readout would be negligible, so that we may assume the latter to thermalize to the bath temperature. Therefore, for a given gas pressure the force noise is in general lower than in the optical case. A stronger bound on CSL could be in principle inferred, at the cost of a much longer measurement time. Second, operation at low coupling does not present the numerous technical issues of an optomechanical coupling. Therefore, a SQUID readout would be still an option if the long term stability of the trap is good enough to resolve a peak with a bandwidth of the order of $1$\,mHz.
	\item The previous considerations hold for a continuous monitoring strategy. For a stroboscopic strategy the readout requires at least two features: (a) a fast and strong position measurement in order to prepare and measure the system in short time; (b) a straightforward procedure for switching on and off the readout without perturbing the system. In this respect, while the feasibility of this scheme is still to be proven, we suggest that a SQUID readout would be unsuitable because of (a), while a tweezer implementation appears more promising than a cavity one because of (b). A more detailed treatment of a realistic implementation of stroboscopic strategy will be the subject of a future paper.
	
\end{enumerate}

\begin{acknowledgments}

The authors acknowledge support from the EU H2020 FET project TEQ (Grant No. 766900). AP has received funding from the European Union’s Horizon 2020 research and innovation programme under the Marie Sklodowska-Curie Grant Agreement No. 749709.

\end{acknowledgments}

\end{document}